\documentclass[aps,prl,preprint,showpacs,showkeys,groupedaddress,nofootinbib]{revtex4-1}
\usepackage{hyperref}
\usepackage[T1]{fontenc}
\usepackage[ansinew]{inputenc}
\usepackage{amsmath}
\usepackage{setspace}
\usepackage{psfrag}
\usepackage{subfigure}
\usepackage{mathtools}
\usepackage{amssymb,lineno,amsfonts}
\usepackage{verbatim}
\usepackage{color}
\usepackage{graphicx}
\usepackage{bm}
\usepackage{dcolumn}
\usepackage[usenames,dvipsnames]{xcolor}
\definecolor{med-blue}{RGB}{25,25,112}
\hypersetup{colorlinks, linkcolor={blue},citecolor={blue}, urlcolor={blue}}
\usepackage[english]{babel}

\newcommand{\bea}{\begin{equation}}
\newcommand{\eea}{\end{equation}}
\newcommand{\ber}{\begin{eqnarray}}
\newcommand{\eer}{\end{eqnarray}}

\begin{document}

%\preprint{}

\title{Nonlocal interactions in a BEC: an Analogue Gravity perspective}


\author{Supratik Sarkar}
\email[]{supratiks@students.iiserpune.ac.in}
%\thanks{}
%\altaffiliation{Indian Institute of Science Education and research, Pune, India}
%\affiliation{University of Maryland, College Park, MD 20742, US}

%\author{Jayanta K Bhattarjee}
%\email[]{jkb@bose.res.in}
%\thanks{}
%\affiliation{S.N. Bose National Centre for Basic Sciences, JD-Block, Saltlake, Kolkata-700098, India}

\author{A. Bhattacharyay}
\email[]{a.bhattacharyay@iiserpune.ac.in}
%\thanks{}
\affiliation{Indian Institute of Science Education and Research, Pune 411008, India}


\date{September 19, 2013}

\begin{abstract}
We add a minimal correction term to the local Gross-Pitaevskii equation to
represent non-locality in the interactions. We show that the effective minimal
non-locality can make the healing length decrease more rapidly with the
increase of $s$-wave scattering length leaving the expression of the velocity of
sound unaltered. We discuss the implication of this result for a Bose-Einstein
Condensate (BEC) being used as an analogue gravity system. The presented result is
important in the context of condensed matter physics as well because one can
considerably change the size of a quantized vortex at finite $s$-wave scattering
length by tuning the healing length.
\end{abstract}

\pacs{03.75.Kk, 03.75.Nt, 04.70.Dy}
\keywords{analogue gravity, BEC, Bogoliubov spectrum, quantized vortex, healing length}

\maketitle

\section{1. Introduction}
Bose-Einstein Condensate (BEC) is considered as an ideal analogue gravity system \cite{barc1} for various reasons. The nano-Kelvin temperature, quantum coherence over large length and time scales, low speed of sound waves, amenability to accurate experimental control and manipulations \cite{zoll1,zoll2} etc make the BEC one of the ideal systems to simulate quantum blackhole geometries and Hawking radiation. The small amplitude collective modes (phonons) in a BEC can see a Lorentzian metric depending on a suitable background Newtonian bulk flow of the condensate. Under suitable conditions, such a barotropic, inviscid, irrotational background flow field can generate an ergo-region for the phonons which cannot overflow the boundary where the bulk velocity exceeds the velocity of sound. It was Unruh's 1981 paper \cite{unruh} which practically started this field of analogue gravity by proposing the possibility of observing Hawking radiation in analogue systems. Following Unruh's seminal paper, a lot of attention has been attracted by the field not only in the connection with the classic trans-planckian problem \cite{jacob} but also for a wide spectrum of issues elegantly compiled in the review by Barcel\'o, Liberati and Visser \cite{barc2}.
\par
BEC of interacting Bosons is explained on the basis of the Bogoliubov theory. Bogoliubov theory identifies the ground state of the interacting BEC to be the vacuum of phonon-like collective modes (excitations) which obey the famous Bogoliubov dispersion relation $\epsilon_p = \left[ \frac{gn}{m}{p}^2+\left(\frac{{p}^2}{2m} \right)^2 \right ]^{1/2}$ where ${p}$ is the momentum of the excitation, $g$ is the approximate two-body interaction strength between bosons, $n$ is the density of the condensate and $m$ is the mass of a single boson. At very small $p=\hbar k$, the dispersion relation becomes that of phonon's i.e. $\epsilon_p = \hbar\omega = cp$ where $c=\sqrt{gn/m}$ is the velocity of sound in the condensate. The small amplitude collective oscillations in a BEC can see a curved spacetime over a moving BEC with velocity $\bf v$ for which the dispersion relation deviates from the exact Bogoliubov dispersion law in the following manner \cite{barc2}
\ber
\epsilon_p=\hbar v^ik_i \pm \left[ \frac{gn}{m}{\hbar k}^2+\left(\frac{{\hbar^2 k^2}}{2m} \right)^2 \right ]^{1/2},
\eer 
where $i$ stands for $x,y,z$ and repeated index implies a sum. In the simplest scenario, the low energy excitations (small $k$), having a linear dispersion relation, respond to the curved spacetime geometry and sees the sonic horizon (where the bulk velocity becomes supersonic) from which an analogue Hawking radiation is expected. 
\par
At small $k$, one can ignore the under-root quartic term in $k$ in Eq.1 and deal with a linear dispersion relation, but, the approximation breaks down at the horizon where ${\bf v} \rightarrow c$ and the quartic term becomes relevant. The presence of the quartic term at the horizon can of course give rise to interesting physics. Some of the works done by Parentani and
coworkers considering a novel density-density correlation probe \cite{seven}, studying hydrodynamic flow over several length scales giving Plankian spectrum with a surface gravity independent
temperature \cite{eight}, considering high frequency dispersion over a wider set of conditions \cite{nine}, are
quite important. Another recent work by Fleurov \emph{et al} taking into account the \textcolor{red}{quantum potential}
term, which is normally neglected in the hydrodynamics of the quantum fluids to simplify the
scenario, shows a second characteristic length scale which is an important observation \cite{ten}. So, the presence of dispersion makes the physics richer, but at the same time, a control over
this dispersion can be of immense importance to look into the system more closely and for
having the group velocity `tunable'. There can be small k problems due to the finite size of the
horizon introducing a large wavelength cut off \cite{zoll1} beyond which sound waves would get just
diffracted. So, (a) controlled access to small wavelength excitations and (b) the finite (small)
velocity of sound are two essential ingredients for the BEC to be used as an analogue system.
In this paper, we discuss the scenario where a minimal correction in the local Gross-Pitaevskii
(GP) equation arising from the structureless non-locality in the interactions decreases the
healing length in such a way, that the quartic term in the dispersion relation can be made very
small on tuning the $s$-wave scattering length keeping the velocity of sound the same as that
of the local GP equation. The present analysis indicates a simple way of achieving both the
essential ingredients mentioned above by the tuning of $s$-wave scattering length for repulsive
interactions. The variation of scattering length in BEC used as an analogue system has been
discussed in \cite{barc3} in a different context.
\par
Considering the same set of basic conditions, as is there in the Bogoliubov theory, namely, (a) the macroscopic occupation of the BEC ground state and (b) diluteness of the condensate, Gross and Pitaevskii derived the famous equation for the complex order parameter of the inhomogeneous condensate known as Gross-Pitaevskii (GP) equation \cite{pita}. GP equation not only correctly reproduces the quantum mechanical Bogoliubov spectrum, its extremely useful in getting explicit density variations of the inhomogeneous condensate in closed form expressions. The GP equation with local interactions is of the form
\bea
i\hbar\frac{\partial}{\partial t}\psi({\bf r},t) = \left ( -\frac{\hbar^2\nabla^2}{2m} + V_{\text{ext}}({\bf r},t) + g|\psi({\bf r},t)|^2  \right )\psi({\bf r},t),
\eea
where $\psi $ is the order parameter of BEC and is related to the condensate density by $n=|\psi({\bf r},t)|^2$, $V_{\text{ext}}$ is the external potential. The GP equation also admits vortex solutions which are particle like (dispersion relation is not linear like that of massless phonons) and are higher energy states stabilized by rotating the the BEC \cite{pita}.
\par
As is already mentioned, the GP equation in its above mentioned form relies on the diluteness condition to keep the nonlinear term local. The diluteness condition reads as $|a| << n^{-1/3}$ i.e. the $s$-wave scattering length is much smaller than the average range of separation between particles in the BEC. The $s$-wave scattering length is given by the relation $g=\frac{4\pi\hbar^2a}{m}$ following Born approximation and has to be kept positive, in the simplest scenario, to keep the BEC ground state thermodynamically stable. On the basis of the above mentioned assumptions, the nonlocal interaction term $\psi({\bf r},t)\int {d{\bf r^\prime}\psi^*({\bf r^\prime},t)V({\bf r^{\prime}-{\bf r}})\psi({\bf r^\prime},t)}$ of the general GP equation is written as $\psi({\bf r},t)\int{d{\bf r^\prime}V({\bf r^{\prime}-{\bf r}})|\psi({\bf r^\prime},t)|^2}=g\psi({\bf r},t)\int{d{\bf r^\prime}\delta({\bf r^{\prime}-{\bf r}})|\psi({\bf r^\prime},t)|^2}=g|\psi({\bf r},t)|^2\psi({\bf r},t)$. The local GP equation can be shown to be derived variationally by minimizing a free energy functional. The asymptotic exactness of the GP energy functional at the dilute limit with two-body interactions was rigorously shown by Lieb \emph{et al} \cite{lieb1,lieb2}. In this proof, at the thermodynamic limit $N \rightarrow \infty$, $a\rightarrow 0$ to keep $Na$ finite and this limit corresponds well with $V({\bf r^\prime}-{\bf r})$ being replaced by a delta function.
%For the $s$-wave scattering length to be positive, in Bogoliubov-treatment of the problem, the actual inter-particle interaction potential is replaced by an effective flat repulsive potential to get $g=\int{d{\bf r}V_{eff}}$. It is argued in this context that, the actual form of the interaction potential is not that important for low energy scattering in the dilute limit, if it produces the right $s$-wave scattering length \cite{pita}. 
\par
Varied nonlocal forms of the interaction potential have been considered to capture new solutions not admitted by local GP theory \cite{cue, shch}. There is a symmetry based recent analysis of nonlocality in GP equation in \cite{lisok}. With the help of nonlocal nonlinearities, roton minimum in the dispersion relation has been predicted which is a typical characteristic of superfluids \cite{santos, ivas, ryan}. An interesting work \cite{deco} considers non-locality in the interactions in the presence of a particular periodic global potential to show that, although the GP free energy and the corresponding solutions show asymptotic correspondence as one moves from the nonlocal to the local limit, but, the asymptotic correspondence in the stability of the solutions is not present \cite{curt}. In \cite{ander}, the effect of quantum fluctuations on the mean field model is treated semi-classically and the ref.\cite{perez} presents an interesting analysis considering nonlocal interactions preventing collapse of the condensate. The work of Andreev \emph{et al} \cite{andr1} on degenerate boson-fermion system derives a similar dispersion relation based on a third order correction as is there in this paper but focuses on different issues \cite{andr2,andr3,andr4}. Rosanov \emph{et al} have done an analysis on internal oscillations of solitons with nonlocal nonlinearity on a similar system that considers a correction term similar to ours with an attractive interaction at large distances \cite{rosa}. There is a nice review article by Yukalov \cite{yuka} addressing theoretical challenges of BEC theory covering nonlocal and disordered potentials.

\section{2. Model}
The general form of the GP equation is 
%\begin{widetext}
\bea
 i\hbar\frac{\partial}{\partial t}\psi({\bf r},t) = \left ( -\frac{\hbar^2\nabla^2}{2m} + V_{ext}({\bf r},t) \right )\psi({\bf r},t)+ \left (\int {d{\bf r^\prime}\psi^*({\bf r^\prime},t)V({\bf r^{\prime}-{\bf r}})\psi({\bf r^\prime},t)} \right )\psi({\bf r},t),
\eea
%\end{widetext}
which one gets by replacing quantum mechanical field (creation/annihilation) operators $\hat{\psi^\dagger}/\hat{\psi}$ by complex numbers (order parameters) $\psi^*/\psi$ in the Heisenberg equations for the fields, where the order parameter is related to the condensate density as $|\psi(\textbf{r},t)|^2=n$. One can do this approximation by considering a macroscopic occupation of the BEC where the non-commutativity of the creation and annihilation operators can be ignored at the mean filed level as is considered in the Bogoliubov theory. 
\par
If one Taylor expands $\psi({\bf r^\prime},t)$ about the position ${\bf r}^{\prime}={\bf r}$ in 1-d by keeping in mind the fact that the interaction potential is symmetric, i.e., $V({\bf r^{\prime}-{\bf r}})\equiv V_{\text{eff}}(x^{\prime} -x)=\frac{g}{\sqrt{2\pi}a}e^{-\frac{|x^{\prime}-x|^{2}}{2a^{2}}}$, one generates the nonlocal GP equation for 1-d inhomogeneities with additional nonlinear term(s) to the conventional local GP equation. The minimal nonlocal GP model for a nonuniform BEC in 1-d is given by, 
%\begin{widetext}
\bea
i\hbar\frac{\partial}{\partial t}\psi({\bf r},t) = \left ( -\frac{\hbar^2}{2m}\frac{\partial ^2}{\partial x^2} + V_{\text{ext}}({\bf r},t) + g|\psi({\bf r},t)|^2\right )\psi({\bf r},t) + \boxed{\frac{1}{6}a^2g\psi({\bf r},t)\frac{\partial ^2}{\partial x^2}|\psi({\bf r},t)|^2 }.
\eea
%\end{widetext}
The above equation effectively considers that, the $\psi({\bf r},t)$ is uniform on the $y-z$ plane (for the sake of simplicity) and also that, there are no small amplitude global modes with spherical or circular symmetry in the system for the unavoidable spatial dependences of the amplitude of such modes.
The gradient term in the Taylor expansion does not contribute due to the symmetry of the interaction and we have kept the next higher order term here. The interaction
potential $V(\textbf{r}^{\prime}-\textbf{r})$ is taken here to be the same flat repulsive potential 
$V_{\text{eff}}(\textbf{r}^{\prime}-\textbf{r})\equiv g\delta(\textbf{r}^{\prime}-\textbf{r})$ considered in the
derivation of the conventional local GP equation \cite{pita}. The $V_{\text{eff}}$ has a range
effectively equal to the $s$-wave scattering length $a$ which is the effective range of interaction
seen by particles undergoing so-called zero-energy scattering. The spherical symmetry of
the interaction is evident from the consideration of $s$-wave scattering only. It is interesting
to note that the first nonlinear term of our expansion is the same as the nonlinear term of
the conventional local GP equation. But, not imposing the drastic $\delta$-function approximation
allows for the appearance of the new term(s) which would have been missing had we applied the $\delta$-function approximation in the beginning.

%In the new term perturbing the system (which is a true perturbation for $a_0$ small), we have considered $\int{d{\bf r}({ x^\prime}-{x})^2V({\bf r^\prime}-{\bf r})}=a_0^2g$, where $a_0$ captures the approximate length scale of the interaction when one puts the interaction strength in terms of $g$ following standard procedure of the Bogoliubov prescription. To make things even simple, in what follows, we would consider $a_0\simeq a$ to reduce number of parameters and would consider the $s$-wave scattering length to roughly represent the range of the nonlocal interactions. Effectively, here, we are considering a somewhat flat potential $V_{eff}$ spread over the radius $\frac{a}{2}$. 
\par
We are not considering the interaction potential to be a $\delta$-function. We are considering $a$ as finite and that legitimizes the inclusion of the correction (nonlinear) term with a second order derivative in the above equation (4). Since, there already exists a term with a second order spatial derivative in the local GP equation, the slowness of density variation over space cannot prevent the additional term from appearing in it. The slowness of the spatial variation of the density is considered here to help drop the higher order terms of the expansion. We would like to emphasize the fact that, this is the first minimal correction term that could be added, keeping the GP dynamics local, in order to incorporate the effect of non-locality of the interactions on top of the $\delta$-function approximation. At the limit $a\rightarrow 0$ (corresponding to the limit taken by Lieb \emph{et al}), this additional term will obviously disappear and we get back the traditional local GP equation (i.e. Eq.(5.2) of \cite{pita}). Apart from relaxing the width of the $\delta$-potential, in all other respect, we are following the standard assumptions of the local GP theory. Moreover, our model is also a local one where the extra term brings in the effect of non-locality to some extent.  
\par
Let us identify a few properties of the equation (4). Like the usual local GP equation, equation (4) also preserves the continuity equation $\frac{\partial n({\bf r},t)}{\partial t}=-\nabla \cdot {\bf j}({\bf r},t)$ where the current density ${\bf j}({\bf r},t) = -\frac{i\hbar}{2m}\Big(\psi^*({\bf r},t)\nabla\psi ({\bf r},t) - \psi({\bf r},t)\nabla\psi^*({\bf r},t)\Big)$. The total mass of the condensate is conserved even after the perturbation is added. Second most important feature preserved by this equation is that, the spatially uniform but oscillating ground state solution $\psi_0=\sqrt{n}e^{-\frac{i\mu t}{\hbar}}$ of the local GP equation (i.e., equation (2)) is still a solution of the system, although, we cannot call it the ground state solution because the concept of the GP free energy no longer applies. Actually, it is well-known that the GP ground state is,
in reality, a metastable state. This conservation of mass is an essential ingredient preserved despite the addition of the term representing the non-locality. The condition of the dynamical
stability of the so-called uniform ground state would remain the same in this modified scenario,
however, this is not that important in the analogue gravity context as the dynamical stability
of a state with an underlying supersonic velocity.

\section{3. Small Amplitude Collective Modes in a nonlocal BEC}
Let us consider a condensate without any confining external (trapping) potential and drop the $V_{\text{ext}}$ term from equation (4). To look at 1-d small amplitude modes, we perturb the uniform oscillatory solution as $ \psi(x,t) = \Big(\psi_0(x)+ \sum_{\substack{j}} \left[u_j(x)e^{-i\omega_j t}+v_j^*(x)e^{i\omega_j t} \right] \Big)e^{-\frac{i\mu t}{\hbar}} $. Putting this expression in equation (4) and linearizing, one can get the dynamics of the amplitude of the perturbations as

\begin{subequations}\label{eq:five}
\ber \label{eq:a}
\hbar\omega_j u_j &=& gnu_j + \left(\frac{a^2gn}{6}-\frac{\hbar^2}{2m} \right)u_j^{\prime\prime} + gnv_j + \frac{a^2gn}{6}v_j^{\prime\prime} \\ \label{eq:b}
-\hbar\omega_j v_j &=& gnv_j + \left(\frac{a^2gn}{6}-\frac{\hbar^2}{2m}\right)v_j^{\prime\prime}+ gnu_j + \frac{a^2gn}{6}u_j^{\prime\prime} \hspace{0.5cm},
\eer
\end{subequations}

where we have used $\mu = gn$ which one gets by putting $\psi(x,t)\equiv\psi_0(x)e^{-\frac{i\mu t}{\hbar}}$ in equation (4) for a uniform gas and in the absence of trapping potential ($V_{\text{ext}}=0$). Considering the ansatz $u_j(x)=ue^{ikx}$ and $v_j(x)=ve^{ikx}$, we get the dispersion relation as
\bea
\hbar^2\omega^2=\frac{\hbar^2k^2gn}{m}+\left(\frac{\hbar^4}{4m^2} - \frac{\hbar^2a^2gn}{6m}  \right)k^4.
\eea  
This dispersion relation immediately tells us that, upto the limit $\frac{4\pi}{3}a^3=\frac{1}{2n}$, the qualitative nature of the dispersion relation remains the same as that of the Bogoliubov spectrum. The excitation energy, however, reduces at larger $k$ compared to the Bogoliubov spectrum as $a$ approaches its limiting value $(3/8\pi n)^{1/3}$. The possibility of keeping the dispersion relation linear by tuning the $s$-wave scattering length is the point of interest here from the perspective of BEC as an analogue gravity system. For an $a$ larger than $(3/8\pi n)^{1/3}$ the sign of the coefficient of the quartic term in $k$ would be negative and more than two-body interactions might be important at this stage if superposition of two-body interactions practically fails to accommodate the physics. The bending of the dispersion curve at long-range interactions shows some tendency to the roton minimum, but, in the present context we are confined to 1-d and do not take into account the structures of the interaction potential which stabilizes excitations and brings the dispersion curve back upwards again. However, the indication that a long range interaction is a generic reason for the creation of another minimum for small wavelength excitations is present here.

\subsection{3.1 \underline{Healing length}}
Let us have a look at the change in the healing length $\xi_0$ which is instrumental in demarcating the phonon length scales from the particle like excitations. The Bogoliubov spectrum is given by
\bea
\hbar^2\omega^2=\frac{\hbar^2k^2gn}{m}+\frac{\hbar^4k^4}{4m^2}.
\eea
At small $k$, this dispersion relation takes the form of a phonon dispersion relation $\hbar\omega=pc$, where $p=\hbar k$. The healing length indicates the point of transition from the phonon spectrum to the particle spectrum where one considers the kinetic and the potential energy balance as $\frac{p^2}{2m}=\frac{\hbar^2}{2m\xi_0^2}\simeq gn$. This relation gives a healing length $\xi_0 = \frac{\hbar}{\sqrt{2mgn}}=\frac{\hbar}{\sqrt{2}mc}=\frac{1}{\sqrt{8\pi an}}$. The relationship from which the healing length is derived, being a balance between the kinetic and the potential energy of excitations, indicates that excitations with a smaller length scale than the healing length are treated as particles (wave packets with mass). Phonon-like excitations typically have wavelengths larger than the healing length. In the modified model, to find the healing length, we have to consider a balance between the quadratic and the quartic terms in the wave number on the right hand side of equation (6) considering
\bea
p^2 c^2= \left (\frac{1}{4m^2}-\frac{a^2gn}{6m\hbar^2}\right )p^4,
\eea 
we get
\bea
\boxed{\xi = \xi_0\left(\frac{1}{2} -  \frac{4}{3}\pi a^{3}n \right)^{1/2}} \hspace{0.3cm}\because \xi_0=\frac{1}{\sqrt{8\pi an}} \hspace{0.2cm}.
\eea 

So, the healing length ($\xi$) decreases with the increase in $a$ which is a much rapid decrease for a constant $n$ than the $a^{-1/2}$ scale of decrease obtained from the conventional local GP model. In fact, where the healing length ($\xi_{0}$) scaling as $a^{-1/2}$ at a constant $n$ becomes zero at $a\to \infty$, but here it becomes zero at an $a \sim n^{-1/3}$; i.e., at a finite $s$-wave scattering length, say some $\varsigma = (8\pi n/3)^{-1/3}$,
\bea \nonumber
\lim_{a \to \varsigma}\hspace{0.1cm}\xi = 0.
\eea 
In Fig.1, we compare the variation of healing length with $a$ for both local ($\xi_{0}$) and nonlocal ($\xi$) cases at various densities of the condensate. The equivalent way of looking at the present scenario is an increase of the effective mass of the particles with an enhancement of the scattering length. Its the Laplacian term appearing as the minimal correction with an opposite sign to the kinetic energy term that renormalizes the effective mass. Interesting to note that, this increase of the effective mass of the particles is not affecting the expression of the velocity of the sound wave in the BEC which comes from the interactions itself. Its the $\delta$-function (local) interaction that fixes the velocity of sound just as in the local GP equation. Thus, the velocity of sound scales as $\sqrt{a}$ at a constant $n$ and this finite velocity of sound at a vanishing of the healing length ($\xi \to 0$) is an interesting situation for an analogue system.
\begin{figure}[h]
%\begin{center}
\includegraphics[angle=-90,scale=0.5]{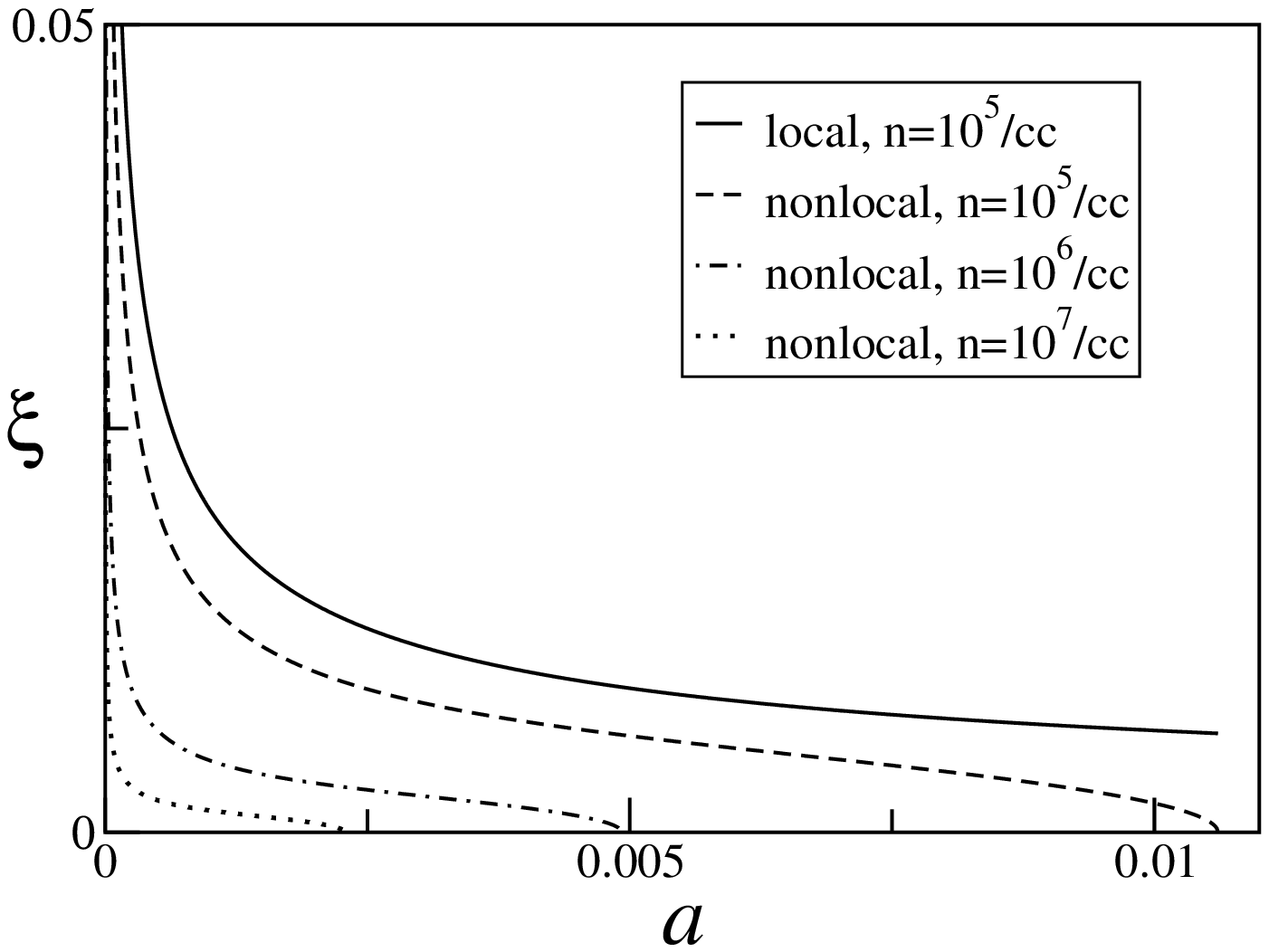}
\caption{\small{Comparison of the changes of healing length for local ($\xi_{0}$) versus nonlocal ($\xi$) BEC with the change of $s$-wave scattering length $a$ at various concentrations ($n$).}}
%\end{center}
\end{figure}
\subsection{3.2 \underline{Trapped BEC}}
The $s$-wave scattering length $a$ can be tuned practically from $-\infty$ to $\infty$ \cite{corn} using Feshbach resonance in a trapped gas. In an asymmetric cigar shaped trapped gas, one gets hydrodynamic modes when the Thomas-Fermi condition $Na/a_{\text{ho}}>>1$ is satisfied. This limit enables one to drop the \textcolor{red}{quantum potential} term (kinetic energy term) in the GP equation where $N$ is the number of particles in the system and $a_{\text{ho}}=\sqrt{\hbar/(m\omega_{\text{ho}})}$ is the new length scale introduced by the harmonic trap \cite{pita}. The additional conditions for such modes to exist are $kL>>1$ and $k\xi << 1$ where $L$ is the size of the condensate. The first of the above two additional conditions requires the wavelength of the phonon be much smaller than the system size and the second one requires that the wavelength of the phonon be much larger than the healing length. This is somewhat a delicate situation posing an upper and lower length scale cut offs.
\par
Let us again have a look to our modified GP equation with the interaction term appearing from the nonlocality rearranged in a convenient way
%\begin{widetext}
\ber \nonumber
i\hbar\frac{\partial}{\partial t}\psi({\bf r},t) = \left ( -\frac{\hbar^2}{2m}\frac{\partial ^2}{\partial x^2}\psi({\bf r},t)+ \frac{1}{6}a^2g\psi({\bf r},t)\frac{\partial ^2}{\partial x^2}|\psi({\bf r},t)|^2\right )+\Big(V_{\text{ext}}({\bf r},t) + g|\psi({\bf r},t)|^2\Big)\psi({\bf r},t)  . \\ \nonumber
\eer
%\end{widetext}
We are using the same 1-d equation considering the modes to reside near the middle of a long cigar-shaped condensate where the harmonic trap, i.e. $V_{\text{ext}}=\frac{1}{2}V_{0}\left(\frac{x}{\eta}\right)^{2}$ with $\eta$ being the range parameter for the external trap, is asymmetric in one of the three possible directions. The first term in the parenthesis on the r.h.s is the effective kinetic energy term which has two Laplacian terms of opposite signs and this sign-difference is at the origin of the `renormalization' of the effective mass term in equation (8) and should work the same way as in the free condensate discussed above. Now, one can talk about tuning the $s$-wave scattering length $a$ by Feshbach resonance and make the healing length $\xi$ as small as possible attaining the Thomas-Fermi condition even for a dilute BEC and can work in the hydrodynamic limit where the local speed of sound is still given by $c_{s}=\sqrt{gn/m}$. Attaining the Thomas-Fermi limit by shrinking $xi$ would also help satisfy the relations $kL>>1$ and $k\xi << 1$ in relatively small systems. Note again that, the $c_{s}$ does not change because it comes from the local interaction term while the nonlocal correction term in the conventional GP equation only affects the kinetic energy sector.

\section{4. BEC as an analogue gravity system}
Considering a single particle state of the form $\psi({\bf r},t)=\sqrt{n({\bf r},t)}e^{\frac{i\theta({\bf r},t)}{\hbar}}$, one can write the the local GP equation in the form
\ber
&&\frac{\partial n}{\partial t} +\nabla . (n{\bf v}) = 0\\
&& \frac{\partial \theta}{\partial t} + \frac{(\nabla\theta)^2}{2m} + V_{\text{ext}} +gn -\frac{\hbar^2}{2m}\frac{\nabla^2\sqrt{n}}{\sqrt{n}}=0
\eer
where ${\bf v}(x,t) =\frac{\partial\theta(x,t)}{\partial x}/m$. If we consider the additional term that we have added to the local GP equation and linearize the above two equations for small amplitude density and phase perturbations as $n \rightarrow n + \hat{n}_1$ and $\theta \rightarrow \theta +\hat{\theta}_1$, we get the following two coupled equations \cite{barc2} for $\hat{n}_1$ and $\hat{\theta}_1$ in the absence of $V_{\text{ext}}$ as
\ber
\frac{\partial \hat{n}_1}{\partial t}+\frac{1}{m}\nabla . \Big(\hat{n}_1\nabla \theta + n\nabla\hat{\theta}_1\Big)=0,\\
\frac{\partial \hat{\theta}_1}{\partial t} + \frac{\nabla\theta . \nabla\hat{\theta}_1}{m} + g\hat{n}_1 - \frac{\hbar^2}{2m}D_2 \hspace{0.06cm}\hat{n}_1=0.
\eer
The term $D_2\hspace{0.06cm}\hat{n}_1$ in 1-d will have the form 
%\begin{widetext}
\bea
D_2\hspace{0.06cm}\hat{n}_1=-\frac{\hat{n}_1}{2}n^{-3/2}\frac{\partial^2n^{1/2}}{\partial x^2}+\frac{n^{-1/2}}{2}\frac{\partial^2(n^{-1/2}\hat{n}_1)}{\partial x^2} \boxed{-\frac{gma^2}{3\hbar^2}\frac{\partial^2\hat{n}_1}{\partial x^2}},
\eea 
%\end{widetext}
where the last term on the r.h.s is coming from the correction term added to the local GP equation.
\par
The dispersion relation in 1-d corresponding to the modified GP equation in the present context would look like
\bea
\boxed{\omega= |\textbf{v}|\hspace{0.04cm}k \pm \left[ \frac{gn}{m}k^2+\left (\frac{\hbar^2}{4m^2} - \frac{a^2gn}{6m}  \right )k^4 \right ]^{1/2}}.
\eea
As has been mentioned already, the most important result of this minimal generalization of the local GP equation, which can be achieved by tuning the $s$-wave scattering length $a$, is that the speed of sound $c_{s}$ is not affected but one can get rid of the under-root quartic term in the dispersion relation. So, in principle, one can go to high momentum scales keeping the dispersion relation linear and at the same time keeping $c_{s}$ small. The possibility of tuning the kinetic energy term out helps not only avoid complications at the sonic horizon $c_{s} \sim |{\bf v}| $, but also gives control over moving from massless (linear dispersion relation) to massive (wave packets) excitations and vice versa. Small wave length excitations with linear dispersion are very important when finite size geometrical constraints apply to the sonic horizon. Obviously, keeping up to the third order term in the expansion of the order parameter would impose restrictions on going to small length scales at some point beyond which one should consider other higher order terms. However, the next higher order term in the Taylor expansion of the order parameter would again be canceled by symmetry of interactions and one has to consider an even higher order term, the effect of which should not be felt if one does not go to a much smaller length scale.  
\par
In the high frequency regime of analogue systems, one in general writes the dispersion relation as
\bea
E^2 = m^2c^4 + \hbar^2k^2c^2 + \hbar^2c^2\bigtriangleup(k,K),
\eea
where $K\sim\xi^{-1}$ (e.g. in the present case) is the inverse length scale set by the system \cite{barc2}. The $\xi$ is the healing length, as already mentioned, can be considered the analogue Planck scale of the system. One considers various forms for the last term in the above equation doing an expansion about $k=0$ which may or may not converge. Some examples being $\pm k^3/K$ and $\pm k^4/k^2$ where the positive and the negative sign correspond to so-called super and subluminal scenarios \cite{barc2}. In the superluminal case the group velocity becomes larger than the velocity of sound for small $k$ modes whereas in the subluminal case it can go below the velocity of sound. The mechanism of thermal radiation emitted by the horizon is very different in these two cases. Our dispersion relation (equation (8)) actually shows that the coefficient of the large $k$ part of the spectrum can go through zero providing a means to smoothly move between the sub and superluminal regimes.

\section{5. Discussions}
To conclude, our simple analysis on the basis of a minimal correction to the local GP
equation to incorporate the non-locality might be giving a similar dispersion relation people
have observed previously in other contexts, but, here we identify that, this minimal correction
indicates a tunability of the healing length leaving almost all the other results of the local
GP theory qualitatively the same. Although we are looking at the present result from an
analogue gravity perspective for obvious reasons, it can prove to be extremely important from
a condensed matter perspective as well. The healing length is a measure of the size of a vortex
and being able to change this healing length to a good extent at finite a can generate a whole
lot of possibilities to deal with vortex dynamics. For example, one may think of melting of
a vortex lattice by the shrinking of the healing length and the emergent relaxation of the
system. We hope to discuss similar things in our later communications. Nevertheless, in the context of analogue gravity, the present result indicates a range of tunability of the dispersion.
This important fact was unnoticed in the context of the BEC so far. Thus, the generic large
momentum divergence of the group velocity of wave packets, where undesirable, can be
controlled to a good extent and excitations of small wave number can be accessed.

%\section{References}

\end{document}